\documentclass[prl,longbibliography,twocolumn,superscriptaddress,amscd,amssymb,verbatim]{revtex4-1}

\usepackage{graphicx}
\usepackage{textcomp}
\usepackage[OT1,T1]{fontenc}
\usepackage{color}
\definecolor{red}{rgb}{1,0,0}

\usepackage{epstopdf}

\usepackage{mathtools} 

\newcommand{\CuD}{ZnCu$_3$(OD)$_6$SO$_4$}
\newcommand{\CuH}{ZnCu$_3$(OH)$_6$SO$_4$}
\newcommand{\CufourD}{Cu$_4$(OD)$_6$SO$_4$}

\begin{document}

\title{Instabilities of Spin-Liquid States in a Quantum Kagome Antiferromagnet}
\author{M. Gomil\v sek}
\affiliation{Jo\v{z}ef Stefan Institute, Jamova c.~39, SI-1000 Ljubljana, Slovenia}
\author{M. Klanj\v sek}
\affiliation{Jo\v{z}ef Stefan Institute, Jamova c.~39, SI-1000 Ljubljana, Slovenia}
\author{M. Pregelj}
\affiliation{Jo\v{z}ef Stefan Institute, Jamova c.~39, SI-1000 Ljubljana, Slovenia}
\author{F. C. Coomer}
\affiliation{ISIS Facility, Rutherford Appleton Laboratory, Chilton, Didcot, Oxon OX11 OQX, United Kingdom}
\author{H. Luetkens}
\affiliation{Laboratory for Muon Spin Spectroscopy, Paul Scherrer Institute, CH-5232 Villigen PSI, Switzerland}
\author{O. Zaharko}
\affiliation{Laboratory for Neutron Scattering and Imaging, Paul Scherrer Institute, 
CH-5232 Villigen PSI, Switzerland}
\author{T. Fennell}
\affiliation{Laboratory for Neutron Scattering and Imaging, Paul Scherrer Institute, 
CH-5232 Villigen PSI, Switzerland}
\author{Y. Li}
\affiliation{Department of Physics, Renmin University of China, Beijing 100872, People's Republic of China}
\author{Q. M. Zhang}
\affiliation{Department of Physics, Renmin University of China, Beijing 100872, People's Republic of China}
\author{A. Zorko}
\email{andrej.zorko@ijs.si}
\affiliation{Jo\v{z}ef Stefan Institute, Jamova c.~39, SI-1000 Ljubljana, Slovenia}

\date{\today}
\begin{abstract}
The emergent behavior of spin liquids that are born out of geometrical frustration makes them an intriguing state of matter.  We show that in the quantum kagome antiferromagnet ZnCu$_3$(OH)$_6$SO$_4$ several different correlated, yet fluctuating states exist.  By combining complementary local-probe techniques with neutron scattering, we discover a crossover from a critical regime into a gapless spin-liquid phase with decreasing temperature.  An additional unconventional instability of the latter phase leads to a second, distinct spin-liquid state that is stabilized at the lowest temperatures.  We advance such complex behavior as a feature common to different frustrated quantum magnets.     
\end{abstract}
\maketitle

In recent years, the intense search for quantum spin liquids (SLs) -- collective highly entangled states in which long-range magnetic order is suppressed by quantum fluctuations -- 
has unveiled states with exotic fractional excitations and, frequently, with topological order \cite{balents_spin_2010}.
The case of the geometrically frustrated two-dimensional (2D) Heisenberg quantum kagome antiferromagnet (QKA) is a prime example \cite{balents_spin_2010,sachdev_kagome-and_1992}.
For this model, after numerous theoretical studies proposed a multitude of different ground states, a gapped SL state has been lately advocated \cite{yan_spin-liquid_2011,depenbrock_nature_2012, jiang_identifying_2012}
and also experimentally indicated \cite{fu2015evidence}.
However, its exact nature, as well as its topological classification, remains unclear \cite{depenbrock_nature_2012, jiang_identifying_2012, he2015distinct, punk2014topological, zaletel2015constraints}.
Moreover, fierce competition seems to exist among candidate SL ground states of the QKA \cite{iqbal2013gapless,iqbal2014vanishing}.
Therefore, the true ground state could be very sensitive to perturbations, leaving room for unconventional instabilities such as nontrivial symmetry breaking \citep{clark2013striped} or topological ordering.
Recently, an instability of this kind has indeed been observed experimentally, but on the related 2D triangular lattice \cite{itou2010instability}. 

\begin{figure}[t]
\includegraphics[trim = 0mm 0mm 0mm 0mm, clip, width=1\linewidth]{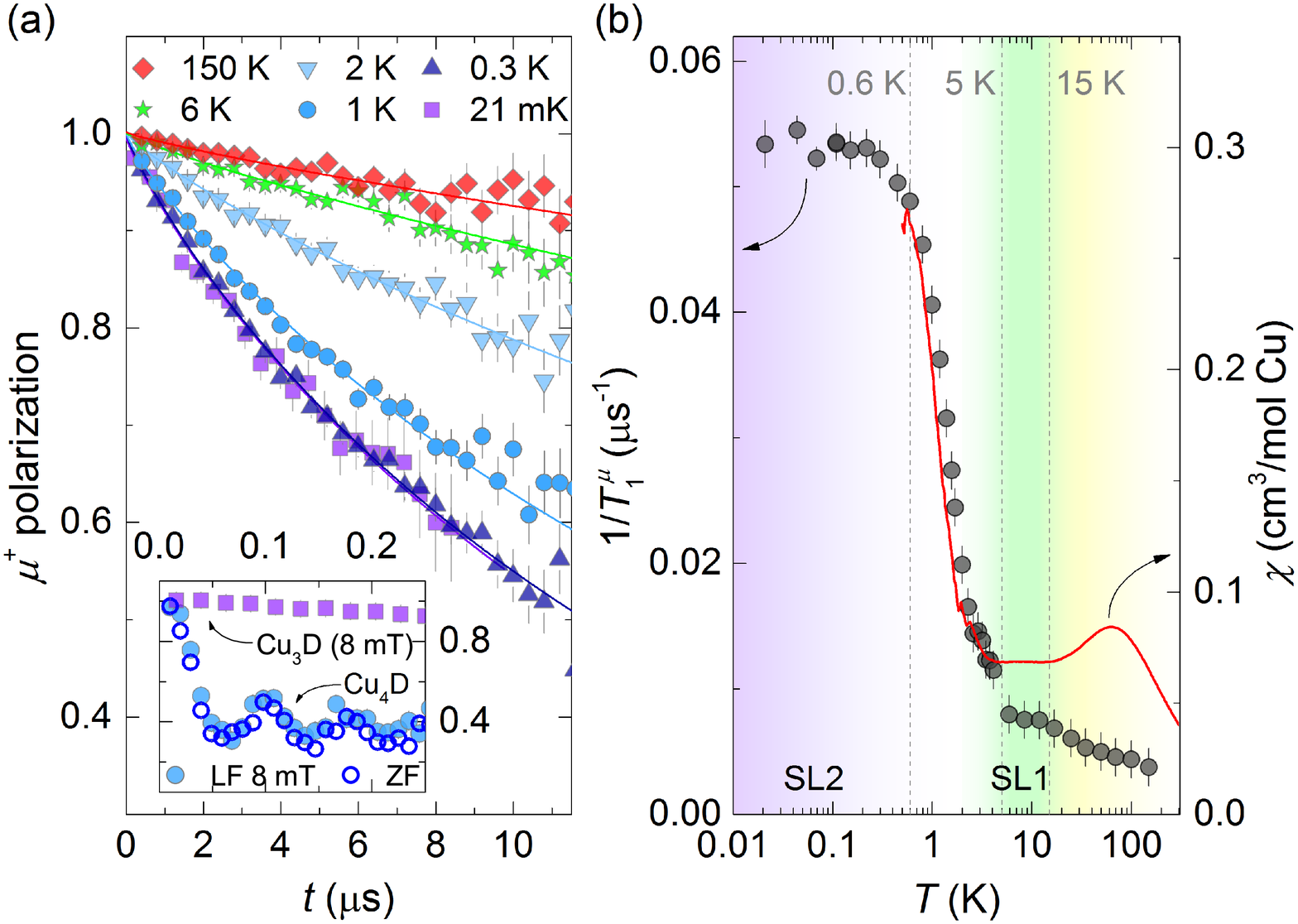}
\caption{(a) The $\mu^+$ polarization in \CuD~(Cu$_3$D) in a LF of 8~mT (symbols) and the corresponding fits from a stretched-exponential model (lines; see text for details).
In the inset the low-$T$ data of Cu$_3$D are compared to 1.6 K data of \CufourD~(Cu$_4$D) in the same LF and in zero field. 
(b) The longitudinal muon relaxation rate $1/T_1^\mu$ compared to the intrinsic kagome susceptibility $\chi_{\rm k}$ measured at 0.1~T \cite{li_gapless_2014}.
The two different SL regimes are denoted by SL1 and SL2.}
\label{fig1}
\end{figure}

To explore possible instabilities of the kagome lattice, we focus on ZnCu$_3$(OH)$_6$SO$_4$, a compound with intriguing magnetic properties \cite{li_gapless_2014} that has recently joined a small number of QKA systems with a possible SL ground state \cite{helton2007spin, mendels2007quantum,faak2012kapellasite, clark2013gapless}. 
Notwithstanding theoretical predictions claiming the contrary, the majority of these systems appears to be gapless \cite{faak2012kapellasite, clark2013gapless,li_gapless_2014}.
Therefore, in-depth experimental studies of new QKA candidates that reveal common features as well as subtle differences among the studied materials are very important.
One particularly relevant issue is the role of impurities in the magnetism of QKAs \cite{dommange2003static, rousochatzakis2009dzyaloshinskii, singh2010valence, kawamura2014quantum, shimokawa2015static}.
This has been experimentally extensively studied in herbertsmithite \cite{de2009scale, han2012fractionalized, nilsen2013low, han2015correlated}, ZnCu$_3$(OH)$_6$Cl$_2$,
which is recognized as the best realization of the QKA model so far, despite substantial (5-10\%) Cu-Zn intersite disorder \cite{mendels_quantum_2010}.
In this compound, such impurities were found to dominate bulk magnetic excitations at low temperatures and low energies ($\lesssim 0.8$~meV) \cite{nilsen2013low, han2015correlated}. 
The compound investigated here closely resembles herbertsmithite in the degree of the Cu-Zn intersite disorder (6 -- 9\%; Ref.~\onlinecite{li_gapless_2014}).
However, the two crystal structures are notably different, as in ZnCu$_3$(OH)$_6$SO$_4$ the Zn$^{2+}$ site lies within the kagome planes, 
while it is positioned between the kagome layers in herbertsmithite.
Consequently, ZnCu$_3$(OH)$_6$SO$_4$ features well isolated 2D kagome planes, where three nonequivalent Cu$^{2+}$ sites are coupled with an average intra-kagome exchange interaction of $J=65$~K \cite{li_gapless_2014}.
Bulk susceptibility ($\chi_{\rm b}$) and heat-capacity ($C_{\rm p}$) measurements failed to detect any sign of magnetic ordering down to temperatures $T$ three orders of magnitude below $J$ \cite{li_gapless_2014}. 
Moreover, a $T$-independent intrinsic kagome susceptibility $\chi_{\rm k}$ [see Fig.~\ref{fig1}(b)] and a linearly increasing magnetic $C_{\rm p}$, which are both characteristic of spinon excitations with a pseudo-Fermi surface, were observed after subtraction of quasi-free-defect contributions from both quantities.  
Surprisingly, these two features signifying a gapless SL state were found in two $T$-regions; between 5 and 15~K (which we call SL1), as well as at $T<0.6$~K (SL2) \cite{li_gapless_2014}. 
Such an intricate, two-state SL behavior, which appears unique among QKA representatives, clearly requires further verification and deeper understanding.

Here we provide a unique experimental viewpoint on the stability of the SL states in ZnCu$_3$(OH)$_6$SO$_4$ that is related to the general issue of the QKA ground state, by 
combining inelastic neutron scattering (INS) with two local-probe spectroscopic techniques employing different coupling mechanisms, nuclear magnetic resonance (NMR) and muon spin relaxation ($\mu$SR).
The temperature- and field-dependent NMR and $\mu$SR spin-lattice relaxation ($1/T_1$) measurements, supported by INS experiments that yield complementary information on the spatial and temporal spin correlations, expose various distinct fluctuating regimes.
$\mu$SR, performed down to 21~mK, gives evidence of a fluctuating SL ground state, while NMR discloses another state with characteristics of a gapless SL at higher temperatures.
In combination with INS, the NMR $1/T_1$ measurements reveal that this high-$T$ SL state is followed by a critical region at even higher temperatures, where the scaling of the dynamical susceptibility with $\omega/T$ is obeyed over several decades in frequency $\omega$.
This complex sequence of correlated electronic states suggesting SL instabilities is likely a more general feature of frustrated antiferromagnets.

$\mu$SR is a particularly powerful technique for detecting even the smallest magnetic fields $B_\mu$, of either nuclear or electronic origin, at the muon stopping site and for quantifying their dynamics  \cite{yaouanc2011muon}.
The measurements were performed on the GPS and LTF instruments, Paul Scherrer Institute (PSI), Switzerland and on the EMU instrument at the ISIS Facility, Ratherford Appleton Laboratory (RAL), United Kingdom. 
The time decay of the muon spin polarization $P(t)$ is driven by the Zeeman coupling of the muon magnetic moment with $B_\mu$. 
It is presented in Fig.~\ref{fig1}(a) for 
a longitudinal field (LF) of 8~mT, which isolates the relaxation due to fluctuating Cu$^{2+}$ electronic moments \cite{sup}.
The background-subtracted \cite{sup} data could be modeled with the stretched-exponential form
$P_{\rm str}(t)={\rm e}^{-(t/T_1^\mu)^{\beta_\mu}}$, with the stretching exponent $\beta_\mu=0.86(6)$.
As $T$ decreases from 150~K, the muon relaxation rate $1/T_1^\mu$ first increases gradually, but then gets substantially enhanced below 5~K [Fig.~\ref{fig1}(b)].
Since the LF is small compared to the frequency $\omega_e$ of fluctuating $B_\mu$ ($\gamma_\mu B_{\rm LF}\ll \omega_{\rm e}\sim k_{\rm B} J/\hbar$; $\gamma_\mu=2\pi\times 135.5$~MHz/T, $k_{\rm B}$ and $\hbar$ denote the muon gyromagnetic ratio, the Boltzmann and the reduced Planck constants, respectively) the standard exponential decay of the field auto-correlation function yields the muon relaxation rate $1/T_1^\mu\propto (\gamma_\mu \Delta B_\mu)^2/\omega_{\rm e}$ \cite{yaouanc2011muon}.
The observed increase of $1/T_1^\mu$ below 5~K is thus a sign of pronounced slowing down of electronic fluctuations and/or enlarged local-field distribution width $\Delta B_\mu$. It coincides with the low-$T$ increase of the intrinsic $\chi_{\rm k}$ [Fig.~\ref{fig1}(b)].
$1/T_1^\mu$ levels off below $\sim$0.6~K [Fig.~\ref{fig1}(b)], signaling persistent dynamics of $B_\mu$ below this temperature. 
The low-$T$ increase of $1/T_1^\mu$ and the relaxation plateau are both features common to different kagome antiferromagnets \citep{mendels2007quantum, kermarrec2011spin, zorko2008easy, faak2012kapellasite, clark2013gapless}.
The unchanged shape of $P(t)$ unquestionably reveals that any kind of electronic freezing is absent in \CuD~at least down to 21~mK, that is $T/J\gtrsim 1/3000$.
This stands in sharp contrast to the related magnetically ordered \CufourD~where an oscillatory $P(t)$ is observed below its N\'eel temperature of 7.5~K \cite{vilminot_nuclear_2006}.
In that compound the first local minimum occurs at $t=\frac{\pi}{\gamma_\mu B_\mu}=0.062$~$\mu$s at 1.6~K [inset in Fig.~\ref{fig1}(a)] and corresponds to a static $B_\mu=60$~mT. 

\begin{figure}[t]
\includegraphics[trim = 0mm 0mm 0mm 0mm, clip, width=0.985\linewidth]{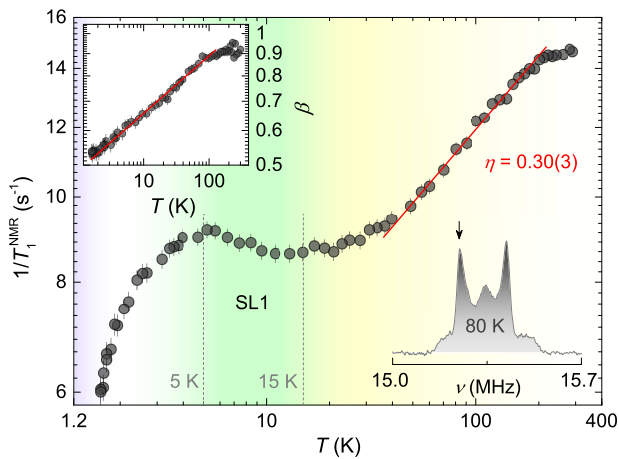}
\caption{The $^2$D NMR relaxation rate $1/T_1^{\rm NMR}$ measured at 2.35~T on the left spectral maximum of \CuD~(arrow in the lower inset, showing the NMR spectrum at 80~K), obtained from the model of magnetic relaxation (see the text). The straight line points out the region where $1/T_1\propto T^{\eta}$. 
The upper inset depicts the $T$-dependence of the stretching exponent $\beta$, also showing a power-law dependence below 100~K.
}
\label{fig2}
\end{figure}

Further insight into the spin fluctuations in \CuD~is provided by $^2$D NMR spin-lattice relaxation. 
These measurements were performed on the lower-frequency spectral maximum (lower inset in Fig.~\ref{fig2}; details on static NMR properties are given in Ref.~\cite{sup}).
The relaxation rate $1/T_1^{\rm NMR}$ is determined by fluctuating local magnetic fields, as the experimental magnetization inversion-recovery curves are explained well with the magnetic-relaxation model for spin-1 nuclei \citep{suter1998mixed}, 
$M(t)=M_0\left[1-\left(1+s\right)\left(\frac{1}{4} {\rm e}^{-(t/T_1)^\beta} + \frac{3}{4} {\rm e}^{-(3t/T_1)^\beta}  \right) \right]$,
with $s<1$ due to imperfect inversion of broad NMR lines and with $\beta$ denoting the stretching exponent. 
$1/T_1^{\rm NMR}$ is set by spin fluctuations at the Larmor frequency $\omega_{\rm L}$,
\begin{equation}
\frac{1}{T_1^{\rm NMR}}\propto \gamma^2 T \sum_{\bf Q}\left|A_{\bf Q} \right|^2\frac{\chi''({\bf Q}, \omega_{\rm L})}{\omega_{\rm L}},
\label{eq1}
\end{equation}
where $\chi''({\bf Q}, \omega_{\rm L})$ stands for the wave-vector dependent dynamical susceptibility and $A_{\bf Q}$ is the coupling between the $^2$D nuclear and Cu$^{2+}$ electronic moments.
In a paramagnetic regime a temperature-independent relaxation rate is expected \cite{moriya1956nuclear}.
In \CuD, deviations from such a behavior are already found at room $T$ and are substantially enhanced below 200~K (Fig.~\ref{fig2}).
The observed decrease of $1/T_1^{\rm NMR}$ is a robust sign of developing spin correlations and is accompanied by a significant increase of the $T_1$-distribution width,
evidenced by $\beta$ decreasing from 0.93(3) at 300~K to 0.53(3) at 1.6~K (upper inset in Fig.~\ref{fig2}). 
Such a $T$-dependent broadening of the $T_1$ distribution is often observed in frustrated antiferromagnets \cite{itou2010instability, shimizu2006emergence, quilliam2011ground, kermarrec2014spin} and is generally recognized as a fingerprint of correlated disordered states.
On the contrary, a trivial reduction of $\beta$ imposed by impurities that lead to a distribution of $A_{\bf Q}$'s is $T$-independent \cite{kermarrec2014spin}.

Between 200 and 40~K, we find $1/T_1^{\rm NMR}\propto T^\eta$ with $\eta=0.30(2)$.
A similar power-law behavior was previously observed in a number of 2D SL candidates, where it was attributed to gapless excitations \cite{shimizu2003spin, itou2008quantum, imai2008cu, olariu200817, dey2013possible}.
$1/T_1^{\rm NMR}\propto T^\eta$ is theoretically expected for a Dirac U(1) SL on the kagome lattice \cite{ran2007projected,imai2008cu}, however, 
for such a state the theory predicts $\chi\propto T$ and $C_{\rm p}^{\rm ZF}\propto T^2$, which both contradict the experimental observations in \CuH~\cite{li_gapless_2014}.
Alternatively, a gapped $Z_2$ SL in the quantum-critical region close to an antiferromagnetically ordered state of a triangular lattice also exhibits power-law NMR relaxation \cite{qi2009dynamics}.
Moreover, $1/T_1^{\rm NMR}\propto T^\eta$ is well established in 1D spin systems \cite{mukhopadhyay2012quantum}.
There, it is considered a hallmark of quantum criticality emerging at $T\lesssim J$ and $\eta$ ranges from positive to negative values as the magnetic field is tuned across the quantum critical region.
With the universality of quantum-critical spin fluctuations  \cite{sachdev1992universal}, the observed power-law dependence in \CuD~may well be a signature of criticality.
In such a case, the energy scale is determined solely by temperature and is independent of the microscopic characteristics of the Hamiltonian. 
Therefore, one expects a scaling of the form
$\chi''(\omega)\propto |\omega|^\nu F\left(\hbar\omega/k_{\rm B}T\right)$ 
\cite{sachdev1992universal}.
\begin{figure}[t]
\includegraphics[trim = 0mm 0mm 0mm 0mm, clip, width=1\linewidth]{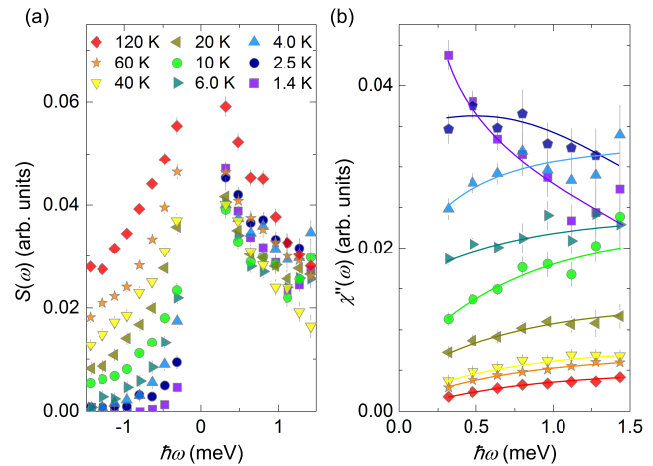}
\caption{(a) The magnetic contribution to the neutron-scattering intensity, $S(\omega)$, in \CuD, where $Q$-integration was performed over wave vectors 0.35~\AA$^{-1}\leq Q\leq\;$1.4~\AA$^{-1}$, and (b) the related dynamical susceptibility $\chi''(\omega)$.
Solid lines are a guide to the eye.
}
\label{fig3}
\end{figure}

\begin{figure}[t]
\includegraphics[trim = 0mm 0mm 0mm 0mm, clip, width=0.985\linewidth]{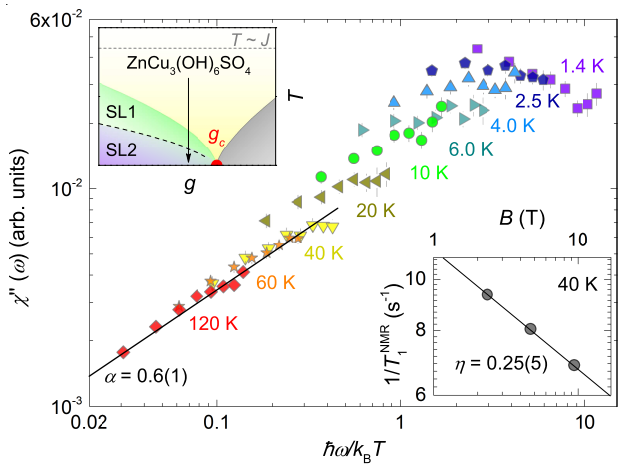}
\caption{Scaling of the dynamical susceptibility with the reduced energy following a power law $\chi''\propto \left(\hbar\omega/k_{\rm B} T \right)^\alpha$ above $\sim$40~K. 
Lower inset: the $1/T_1^{\rm NMR}\propto B^{-\eta}$ field dependence of the $^2$D NMR relaxation rate at 40~K.
Upper inset: a schematic $T-g$ phase diagram with a quantum critical point between SL and some other phase at the critical tuning parameter value $g_c$ and the associated quantum critical region (yellow). 
The SL instability that leads to two distinct SL phases is indicated by the dashed line.
}
\label{fig4}
\end{figure}

In order to verify this interesting possibility in \CuD, we performed powder INS measurements. 
The data were collected on the FOCUS instrument at PSI, Switzerland.
After using an innovative method for the subtraction of background we dwvwlopwd in Ref.~\onlinecite{sup}, the magnetic contribution to the scattering intensity, $S(\omega)$, was obtained by $Q$-integration (0.35~\AA$^{-1}\leq Q\leq\;$1.4~\AA$^{-1}$). 
From $S(\omega)$ the imaginary part of the dynamical susceptibility, $\chi''(\omega)=S(\omega)\left[1-{\rm e}^{-\hbar\omega / k_{\rm B}T}\right]$, was derived (Fig.~\ref{fig3}).
Interestingly, above 40~K, $\chi''(\omega)$ was found to scale with reduced energy as
$\chi''(\omega)\propto (\hbar\omega/k_{\rm B}T)^\alpha$, where $\alpha=0.6(1)$ (Fig.~\ref{fig4}).
Such a scaling law leads, according to Eq.~(\ref{eq1}), to
$1/T_1^{\rm NMR}\propto (T/\omega)^{\eta}$
with $\eta=1-\alpha=0.4(1)$.
This is in good agreement with $\eta=0.30(2)$ obtained from the NMR data.
Moreover, the predicted power-law field dependence $1/T_1^{\rm NMR}\propto \omega^{-\eta}\propto B^{-\eta}$ is also observed experimentally, with $\eta=0.25(5)$ (lower inset in Fig.~\ref{fig4}), thus providing additional verification of the proposed criticality.
We note that similarly $B$-dependent $1/T_1^{\rm NMR}$ is found in quantum-critical 1D systems \cite{mukhopadhyay2012quantum},
while impurities would typically yield a more pronounced dependence, with  $1/T_1^{\rm NMR}\propto B^{-1}$ \citep{klanjvsek2015phonon}. 
The astonishing feature that emerges when combining NMR and INS results on \CuD~in the $T$-range $40 - 200$~K is that the $\chi''(\omega)$ scaling observed in the INS data in the energy range $0.3 - 2$~meV extends to several orders of magnitude lower energies, as $\hbar\omega_{\rm L}=0.06$~$\mu$eV in NMR.
Such a broad range of universal fluctuation indeed appears to be a fingerprint of criticality. 
Scale-free intrinsic kagome-lattice correlations that persist to high $T$ at high energies were also observed in herbertsmithite \cite{de2009scale}. 
There, the impurity contribution to $\chi''(\omega)$ is limited to $T<10$~K \cite{nilsen2013low} and $\hbar\omega<0.8$~meV \cite{han2015correlated}.
As the effective low-$T$ Curie-Weiss temperature of the impurity spins is $\sim\,$-1~K in \CuH~\cite{li_gapless_2014} as well as in herbertsmithite \cite{mendels_quantum_2010} and the amount of impurities in both compounds is very similar, we exclude any role of quasi-free impurities in the scaling presented in Fig.~\ref{fig4}.
On the other hand, any observed scaling at low $T$ and $\omega$ \cite{helton2010dynamic} could be biased by impurity contributions.

Below 40~K, the high-$T$ scaling of the dynamical susceptibility is not valid, as the $\chi''(\omega,T)$ data sets from different temperatures no longer overlap (Fig.~\ref{fig4}).  
At the same temperature, the power-law dependence of $1/T_1^{\rm NMR}$ (Fig.~\ref{fig2}) starts to flatten.
It exhibits a shallow minimum at $\sim$15~K and a maximum at $\sim$5~K (Fig.~\ref{fig2}).
In sharp contrast, $\beta$ evolves monotonically, proving that such behavior of $1/T_1^{\rm NMR}$ is intrinsic.
The $1/T_1^{\rm NMR}(T)$ dependence remains qualitatively unchanged even if we use a fixed $\beta$ to fit the data. 
The essentially $T$-independent region of $1/T_1^{\rm NMR}$ between the shallow minimum and the maximum coincides with the temperature-independent region of intrinsic susceptibility [Fig.~\ref{fig1}(b)] and the linearly-dependent heat capacity \cite{li_gapless_2014}. 
It can thus be attributed to a correlated SL state of the system below $\sim$15~K.
We note that the flattening of $1/T_1^{\rm NMR}$ below 40~K cannot be attributed to weakly-interacting impurities, as the measured NMR-line shift is driven by intrinsic kagome spins at least down to 10~K \cite{sup}, which differs from the situation encountered in herbertsmithite \cite{imai2011local}.

The dynamical properties of the magnetic state in \CuD~change again below 5~K, as a steady decrease of $1/T_1^{\rm NMR}$ is observed once more upon cooling (Fig.~\ref{fig2}).
Such a rich $T$-dependence clearly reflects multiple diverse states that the system goes through with changing temperature.
At the same time, no anomalies are observed in the static properties of the NMR spectra \cite{sup}.
Interestingly, a qualitatively similar behavior of $1/T_1^{\rm NMR}$ with two different power-law correlated regimes and a flat $T$-region in-between  was found in organic triangular-lattice SL candidates \cite{shimizu2006emergence, itou2010instability}.
In EtMe$_3$Sb[Pd(dmit$_2$)]$_2$, such an intricate behavior was attributed to a symmetry breaking and/or a topological instability of a gapless SL phase \cite{itou2010instability}. 
Based on our $1/T_1^{\rm NMR}$ data, we propose that a similar crossover between different SL1 and SL2 phases is at work in \CuD.

This crossover is further witnessed by the muon relaxation rate $1/T_1^{\mu}$ [Fig.~\ref{fig1}(b)], which suddenly starts to increase below 5~K, after exhibiting a very gentle $T$-dependence between 15 and 5~K, being in line with the high-$T$ SL1 phase.
We note that $\mu$SR yields a somewhat different, more indirect viewpoint on the spin dynamics than NMR.
Namely, the $\mu$SR experiment was performed in a small external field ($B_{\rm LF}\ll B_{\mu}$), therefore Eq.~(\ref{eq1}), which is derived using time-dependent perturbation theory in the limit of large external fields, is not directly applicable \cite{lacroix2011introduction}. 
Moreover, in $\mu$SR, contrary to NMR, the long-range dipolar interactions typically overshadow the contact hyperfine interaction in magnetic 
insulators \cite{yaouanc2011muon}.
Therefore, in \CuD, the effective spatial averaging of the dipolar interaction makes the muon relaxation due to the relatively dense impurity spins dominant at low temperatures, where $\chi_{\rm k}\ll\chi_{\rm b}$ \cite{li_gapless_2014}.
The same behavior was indeed observed in herbertsmithite, where the low-$T$ $1/T_1^{\mu}$ scaled linearly with the quantity of Cu$^{2+}$ impurities on the Zn site \cite{kermarrec2011spin}.
The increase of $1/T_1^{\mu}$ below 5~K can in our case then be ascribed to evolving correlations between impurity and intrinsic kagome spins.
Judging from the relaxation plateau, these correlations saturate within the low-$T$ SL2 state below 0.6~K, suggesting that an effective impurity coupling is established.
Because of a finite $T$-independent $\chi_{\rm k}$ and linear $C_{\rm p}(T)$ below 0.6~K \cite{li_gapless_2014} the SL2 state appears to be gapless, with any gap smaller than $J/3000$ as evidenced by persisting muon relaxation observed down to at least 21~mK.

From a theoretical standpoint, a gapped SL ground state, possibly of $Z_2$ type, is expected for the isotropic Heisenberg QKA model  \cite{yan_spin-liquid_2011, jiang_identifying_2012, depenbrock_nature_2012}.
The weight of magnetic excitations could, however, be considerably shifted towards zero energy due to a proximate valence bond solid (VBS) state \cite{punk2014topological}, as was proposed from experimental observations on herbertsmithite \cite{han2012fractionalized}.
Alternatively, quantum criticality between the $Z_2$ SL and an antiferromagnetically ordered state can be induced 
by the Dzyaloshinskii-Moriya interaction \cite{cepas2008quantum, huh2010quantum}, which is known to be sizable in other Cu-based QKA representatives \cite{zorko2008dzyaloshinsky, zorko2013dzyaloshinsky} and could also justify a gapless state \cite{seman2015many}.
Such a magnetic anisotropy that acts as a tuning parameter being slightly off its critical value $g_{\rm c}$, may explain the scaling behavior observed in \CuD~around $T\sim J$ and the selection of the gapless high-$T$ SL1 state around 15~K ($T/J \sim 0.23$), in the sense of the schematic phase diagram shown in Fig.~\ref{fig4}. 
What is intriguing is the crossover from this state into the SL2 state that is stabilized at even lower temperatures, that is below 0.6~K ($T/J\sim0.01$).

With two clearly distinguishable SL regimes the investigated compound is unique among QKAs.
The crossover between the two regimes observed in NMR and $\mu$SR relaxation is not reflected in the static NMR properties.
Hence, the distinction between the two SL states must originate from spin excitations.
Therefore, this crossover may well be of a topological character or due to nontrivial symmetry breaking.
In fact, for gapless SL states with a spinon pseudo-Fermi surface various spinon-pairing instabilities can be expected at low temperatures \cite{clark2013striped,lee2007amperean,galitski2007spin}.
Alternatively, impurities due to the Cu-Zn intersite disorder inherent to the investigated system that are strongly coupled to the kagome lattice could pin spinons at low $T$ and thus provide a spinon-instability mechanism.
While such a premise calls for further theoretical investigations, we note that the intriguing instability of the SL observed in \CuH~appears to be a more general feature of geometrically frustrated antiferromagnets, common to 2D kagome and organic triangular lattices \cite{itou2010instability} that possess a much smaller quantity of impurities, and possibly even to the 3D hyperkagome lattice \cite{zhou20084}.
We expect that such an emergent behavior of SLs is a fingerprint of particular type of low-energy excitations,
common to various geometrically frustrated spin lattices.

\acknowledgments
We thank P. Mendels and F. Bert for fruitful discussions.
The financial support of the Slovenian Research Agency (Program No.~P1-0125) and the Swiss National Science Foundation (SCOPES project IZ73Z0\_152734/1 and Grant No.~200021\_140862) is acknowledged.
Q.~M.~Z. was supported by the NSF of China and the Ministry of Science and Technology of China (973 projects: 2012CB921701).
The $\mu$SR project has received funding from the European Union's Seventh Framework Programme for research, technological development and demonstration under the NMI3-II Grant number 283883.
Neutron scattering experiments were performed at the Swiss spallation neutron source SINQ, Paul Scherrer Institute, Villigen, Switzerland.
We thank C.~Baines for his technical assistance at PSI.

%

\begin{widetext}
\vspace{19cm}
\begin{center}
{\large {\bf Supplementary information:\\

Instabilities of Spin-Liquid States in a Quantum Kagome Antiferromagnet}}\\
\vspace{0.5cm}
M. Gomil\v sek,$^{1}$ M. Klanj\v sek,$^{1}$ M. Pregelj,$^1$ F. C. Coomer,$^{2}$ H. Luetkens,$^{3}$\\
 O. Zaharko,$^{4}$ T. Fennell,$^{4}$ Y. Li,$^{5}$ Q. M. Zhang,$^{5}$ and A. Zorko$^{1,*}$
\vspace{0.3cm}

{\it
$^1$Jo\v{z}ef Stefan Institute, Jamova c.~39, SI-1000 Ljubljana, Slovenia\\
\vspace{0.1cm}
$^2$ISIS Facility, Rutherford Appleton Laboratory, Chilton, Didcot, Oxon OX11 OQX, United Kingdom\\
\vspace{0.1cm}
$^3$Laboratory for Muon Spin Spectroscopy, Paul Scherrer Institute, CH-5232 Villigen PSI, Switzerland\\
\vspace{0.1cm}
$^4$Laboratory for Neutron Scattering and Imaging, Paul Scherrer Institute, 
CH-5232 Villigen PSI, Switzerland\\
\vspace{0.1cm}
$^5$Department of Physics, Renmin University of China, Beijing 100872, People's Republic of China\\
}

\end{center}
\end{widetext}

\section{$\mu$SR field-decoupling experiment}
We used powder samples of \CuD~and references \CuH~and \CufourD, all of the same high quality as previously reported in Ref.~\onlinecite{li_gapless_2014}.
The background signal in the $\mu$SR experiments was determined by comparing the experimental data sets obtained from various instruments with the data sets from the GPS instrument operating in the veto mode.
It was found to vary between $\sim$0 and 30\%, depending on the particular instrument. The GPS and LTF instruments at the Paul Scherrer Institute, Switzerland and the EMU instrument at the ISIS facility, Rutherford Appleton Laboratory, United Kingdom were used.

\begin{figure}[t]
\includegraphics[trim = 0mm 5mm 0mm 0mm, clip, width=1\linewidth]{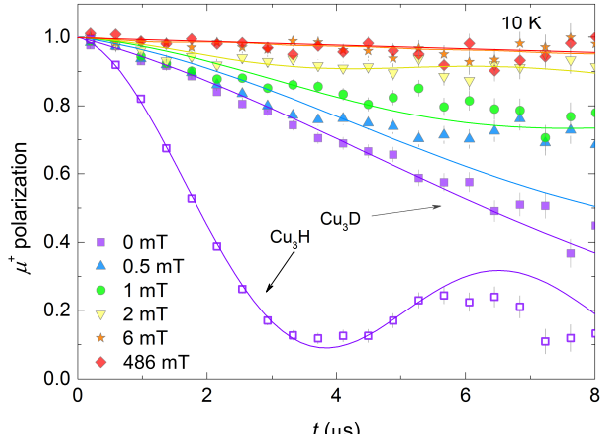}
\caption{
(a) The field dependence of $\mu^+$ polarization in \CuD~(solid symbols) and the zero-field polarization in \CuH~(open symbols) measured at 10 K. 
Solid lines show the simultaneous fit of the model of a OH(D)-$\mu$ complex to the experimental ZF and LF data of both compounds (see text for details).
}
\label{fig1s}
\end{figure}

The time evolution of the $\mu^+$ polarization $P(t)$ in \CuD~is plotted in Fig.~\ref{fig1s} as a function of the applied longitudinal magnetic field (LF) at 10~K. 
This field-decoupling experiment clearly establishes significantly different relaxation functions $P(t)$ of the \CuD~and \CuH~samples in zero applied field (ZF), as well as a notable dependence of $P(t)$ on the applied LF already for fields of only a few mT.
A similar relaxation of the $\mu^+$ polarization in LF was observed before in herbertsmithite \cite{mendels2007quantum} and was attributed to nuclear dipolar fields due to the formation of a OH-$\mu$ complex \cite{schenck1971slow}. 
This model implies oscillations in $P(t)$.
In contrast, in deuterated samples a monotonous $P(t)$ in the experimental time window is expected, because of the reduced nuclear gyromagnetic ratio, $\gamma_{\rm D}/\gamma_{\rm H}=0.154$.
Assuming the same muon stopping site in both \CuH~and \CuD, we performed a simultaneous fit of zero-field relaxation functions

\small 
\begin{widetext}
\begin{subequations}
\begin{align}
P_{\rm H}(t) &=(1-P_{\rm bgd})\frac{1}{6} \left[1+2\cos\left(\frac{1}{2}\omega_{\rm H} t\right)+\cos\left(\omega_{\rm H} t\right)+2\cos\left(\frac{3}{2}\omega_{\rm H} t\right)\right]{\rm e}^{-t/T_1^\mu}+P_{\rm bgd},\\
P_{\rm D}(t) &=(1-P_{\rm bgd})\frac{1}{9} \left[3+2\cos\left(\sqrt{3} \omega_{\rm D} t\right)+\frac{4}{3+\sqrt{3}}\cos\left(\frac{3-\sqrt{3}}{2}\omega_{\rm D} t\right)+\frac{4(2+\sqrt{3})}{3+\sqrt{3}}\cos\left(\frac{3+\sqrt{3}}{2}\omega_{\rm D} t\right)\right]{\rm e}^{-t/T_1^\mu}+P_{\rm bgd},
\end{align}
\end{subequations}
\end{widetext}
\normalsize 
for the hydrogen and deuterium nuclei, respectively, and numerically calculated relaxation functions in finite longitudinal fields \cite{lord2000muon} to the experimental data.
Here, $\omega_{\rm H(D)}=\hbar\frac{\mu_0}{4\pi}\frac{\gamma_{\rm H(D)}\gamma_\mu}{r^3}$.
The fit confirms the model of OH(D)-$\mu$ complex formation (see Fig.~\ref{fig1s}) and yields the muon--hydrogen-isotope distance of $r=1.56$~\AA, which is very similar to the values found in other similar compounds \cite{mendels2007quantum, kermarrec2011spin}.
A small muon relaxation rate due to electronic moments $1/T_1^\mu\sim 0.01$~$\mu$s$^{-1}$ and a background contribution $P_{\rm bgd}=0.17$ are found. 
The LF experiment thus reveals that at 10~K the relaxation of $P(t)$ in small LF is mostly due to small local fields of nuclear origin.
When the LF exceeds 6~mT the muon relaxation is decoupled from the nuclear fields and is solely due to fluctuating electronic moments of Cu$^{2+}$ ions. It remains essentially unaltered under an increase of the LF up to several hundreds of mT.
Therefore, to effectively track the dynamics of the fluctuating Cu$^{2+}$ electronic moments and suppress the relaxation due to nuclear fields within the OH-$\mu$ complex  a field of 8~mT was applied and a deuterated powder sample was used (Fig.~1 in the main text).
\begin{figure}[b]
\includegraphics[trim = 0mm 0mm 0mm 0mm, clip, width=1\linewidth]{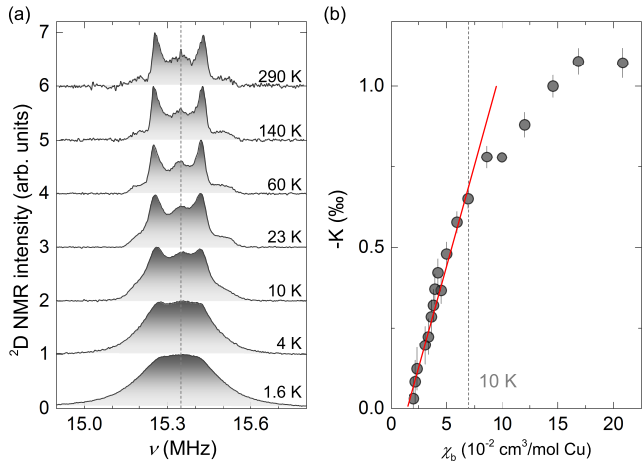}
\caption{
(a) The temperature evolution of the $^2$D NMR spectra of \CuD~measured in 2.35~T.
The spectra are displaced vertically for clarity. 
The vertical line corresponds to the Larmor frequency $\nu_{L}$, i.e. to zero Knight shift. 
(b) Scaling of the Knight shift $K$ with the bulk susceptibility.
The solid line highlights a linear high-temperature dependence.
}
\label{fig2s}
\end{figure}

\section{$^2$D NMR spectra}
 
\begin{figure}[t]
\includegraphics[trim = 0mm 0mm 0mm 0mm, clip, width=1\linewidth]{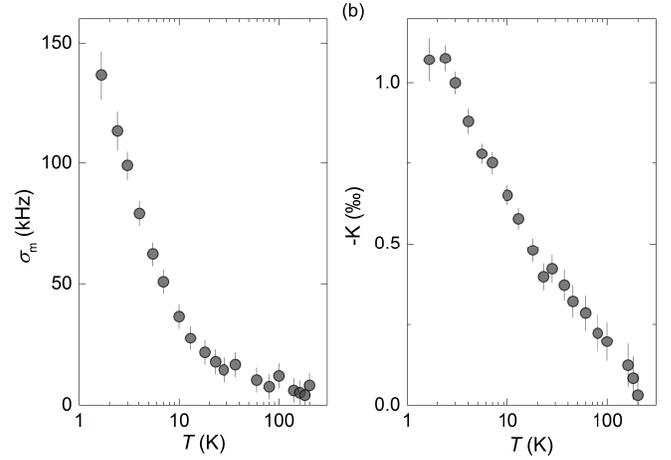}
\caption{
(a) The temperature dependence of (a) the magnetic part of the $^2$D NMR line broadening and (b) Knight shift in \CuD~measured at 2.35~T.
}
\label{fig3s}
\end{figure}

The temperature dependence of $^2$D NMR spectra of \CuD~measured by a frequency sweep in a fixed magnetic field of 2.35~T on a custom-made NMR spectrometer is shown in Fig.~\ref{fig2s}(a). 
The spectra were measured with a solid-echo pulse sequence and a pulse length of 10~$\mu$s.
At high temperatures, a typical quadrupolar powder spectrum -- spin-1 nuclei yield a Pake doublet \cite{abragam1961principles} -- due to the coupling of the nuclear quadrupolar moment with the electric-field gradient is observed.
With decreasing temperature, a spectral broadening and a shift from the Larmor frequency $\nu_{\rm L}$ towards lower frequencies are observed (Fig.~\ref{fig3s}). 
Both are due to magnetic couplings between the nuclear and the electronic magnetic moments.
Notably, no anomalies are observed in either of the two parameters between 5 and 15~K, where the NMR relaxation rate exhibits unusual behavior (see Fig.~2 in the main text). 

The Knight shift 
$K=(M_1-\nu_{\rm L})/\nu_{\rm L}$, 
where $M_1$ denotes the first moment of the NMR line, scales linearly with the bulk susceptibility down to $\sim$10~K, where it starts to saturate [Fig.~\ref{fig2s}(b)].
Such a behavior can be attributed to the presence of impurities in the form of alien Cu$^{2+}$ moments on the Zn sites and Zn$^{2+}$ vacancies on the kagome sites due to Cu-Zn intersite disorder, which was estimated to be 6 -- 9\% \cite{li_gapless_2014}.
In a simple two-component model, one can then separate the Knight shift into two contributions,
$K=A_{\rm k}^{\rm iso}\chi_{\rm k}+A_{\rm i}^{\rm iso}\chi_{\rm i}$, 
where $A_{\rm k}^{\rm iso}$ and $A_{\rm i}^{\rm iso}$ represent the isotropic contact hyperfine coupling constants of the $^2$D nuclear spin with the intrinsic kagome and impurity spins, respectively.
The bending of the $K(\chi_{\rm b})$ curve away from linearity for $\chi_{\rm b} \gtrsim 0.07$~cm$^3$/mol~Cu, i.e. for $T\lesssim 10$~K, where $\chi_{\rm k}\ll\chi_{\rm i}$, reveals that demagnetization effects are small and that $A_{\rm k}^{\rm iso}\gg A_{\rm i}^{\rm iso}$. 
The high-temperature linearity, on the other hand, yields an average coupling constant $A_{\rm k}^{\rm iso}=77$~mT/$\mu_{\rm B}$ of $^2$D nuclei with 
the two neighboring electronic spins on the Cu sites that are exchange-coupled via that particular OD$^-$ bridge.
%

The magnetic part of the spectral broadening was obtained from the second central moment of the NMR line $M_2=\overline{\left(\nu-M_1\right)^2}$, as
$\sigma_{\rm m}=\sqrt{M_2-\sigma_\infty^2}$,
where $\sigma_\infty=M_2(T\rightarrow\infty)$ is related to quadrupolar broadening.
Magnetic broadening reflects a distribution of coupling constants and/or local susceptibilities \cite{kermarrec2014spin},
$\sigma_{\rm m}/\nu_{\rm L}\sim\Delta\left(A\chi\right)$. 
Both are expected in the investigated powder sample, $\Delta A$ due to anisotropic couplings, and $\Delta \chi$, since spin vacancies in the QKA are known to induce extended staggered magnetization profiles in their vicinity \cite{rousochatzakis2009dzyaloshinskii}, with $\Delta\chi/\chi\sim 1$ \cite{poilblanc2010impurity}.

\section{INS background determination}

Since many events, originating from several different neutron-scattering processes, contribute to the observed intensity, it can be very tedious, or even impossible, to extract the intrinsic scattering of a sample. One should disentangle the (possibly tiny) signal of the sample from the background scattering of the sample environment, air scattering and other sources of scattering separate from the sample.
A standard approach is to perform an additional measurement with an empty sample holder (empty can) and assume that this represents the background contribution to the scattering intensity in the measurements with the sample \cite{han2012fractionalized}.
However, this approach has shortcomings due to the multiple scattering processes involved and the finite statistics of empty-can data, which introduce additional uncertainties after the empty-can background subtraction is done.
We instead used a background-determination procedure that generalizes the approach used in Refs.~\onlinecite{helton2007spin,helton2010dynamic} and relies on just two postulates:
a) that the temperature dependence of the background contribution is known (experimentally, it is usually approximately temperature-independent) and b) that the sample is well thermalized, so that we can apply the detailed-balance considerations. No prior assumptions or restriction are imposed on the temperature dependence of the sample contribution, in contrast to the procedure used in Ref.~\onlinecite{de2009scale}.

We consider the following form of the measured total neutron-scattering intensity, 
\begin{widetext}
\begin{equation}
S_{\rm t}(\omega, {\bf Q}, T) = S_{\rm bgd}(\omega, {\bf Q}) + {\left( 1 - e^{-\frac{\hbar\omega}{k_{\rm B}T}} \right)}^{-1} \chi''(\omega, {\bf Q}, T),
\label{inseq1}
\end{equation}
\end{widetext}
where $S_{\rm bgd}$ is a temperature-independent background contribution and $\chi''$ is the imaginary part of the dynamical susceptibility of the sample at the energy transfer $\omega$, wave vector $\bf Q$ and temperature $T$.

The principle of detailed balance implies that the imaginary part of the dynamical susceptibility $\chi''$ is antisymmetric in $\omega$ \cite{chaikin2000principles}, which enables us to exactly extract the background contribution by combining the measured neutron-scattering intensities $S_{\rm t}$ from opposite energy transfers $\pm\omega$ at given ${\bf Q}$ and $T$,
\begin{widetext}
\begin{equation}
S_{\rm t}(-\omega, {\bf Q}, T) - S_{\rm t}(\omega, {\bf Q}, T) e^{-\frac{\hbar\omega}{k_{\rm B}T}} = S_{\rm bgd}(-\omega, {\bf Q}) - S_{\rm bgd}(\omega, {\bf Q}) e^{-\frac{\hbar\omega}{k_{\rm B}T}}.
\label{inseq2}
\end{equation}
\end{widetext}
The only two unknowns in Eq.~(\ref{inseq2}) at fixed $\omega$ and ${\bf Q}$ are the two background contributions $S_{\rm bgd}(\pm\omega, {\bf Q})$. We can obtain these by fitting Eq.~(\ref{inseq2}) to data measured at two or more temperatures $T$. With thus determined background contribution $S_{\rm bgd}$, the dynamical susceptibility of the sample $\chi''$ can be then calculated by Eq.~(\ref{inseq1}).

A strong feature of this method is that ${\bf Q}$-integration and/or powder averaging can be performed either before or after background subtraction, with no change in the result. This is because both Eq.~(\ref{inseq1}) and Eq.~(\ref{inseq2}) are linear in $(S_{\rm bgd}, S_{\rm t}, \chi'')$ and otherwise ${\bf Q}$-independent. Moreover, this background-determination procedure works equally well even if we postulate a different, but known, temperature dependence of the background. The only two requirements are (a) measurements on a symmetric interval of $\omega$ around $\omega = 0$ and (b) data sets at two or more temperatures. To obtain the smallest uncertainty in the results, the input $S_{\rm t}(\omega, {\bf Q}, T)$ should represent raw data with no other background subtraction done beforehand (empty-can or otherwise), since subtracting any finite-statistics temperature-independent data from $S_{\rm t}$ leaves the obtained $\chi''$ unchanged, while it increases the estimated uncertainty. With data at different temperatures, the empty-can measurements thus become redundant, thus saving the measurement time. The empty-can background determination is replaced by an almost equally general background-determination procedure, which is suitable for all types of inelastic neutron scattering measurements, either related to magnetism, lattice vibrations or any other property of the sample.

\begin{figure}[b]
\includegraphics[trim = 0mm 0mm 0mm 0mm, clip, width=1\linewidth]{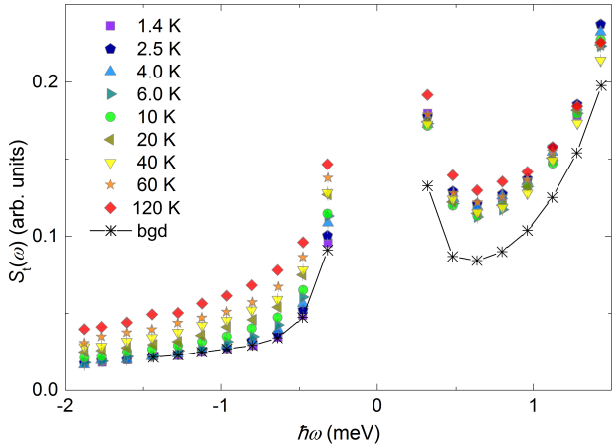}
\caption{The temperature dependence of the total neutron-scattering intensity $S_{\rm t}(\omega)$ in \CuD, integrated over wave vector magnitudes 0.35~\AA$^{-1}\leq Q\leq\;$1.4~\AA$^{-1}$ and the corresponding background contribution $S_{\rm bgd}$.}
\label{fig4s}
\end{figure}

As already mentioned above, this procedure is a generalization of the approach used in Refs.~\onlinecite{helton2007spin,helton2010dynamic}, where a measurement at extremely low temperatures in the millikelvin range was required to obtain $S_{\rm bgd}$ at $\omega<0$ (by equating it with the measured intensity there) and one at higher temperatures to subsequently obtain $S_{\rm bgd}$ at $\omega>0$ (by using the principle of detailed balance and a known background at $\omega<0$). Comparing this to our procedure, we see that the low-$T$ step corresponds to using Eq.~(\ref{inseq2}) in the limit $k_{\rm B}T \ll \hbar\omega$, since then $S_{\rm t}(-\omega, {\bf Q}, T) \simeq S_{\rm bgd}(-\omega, {\bf Q})$, and the high-$T$ step corresponds to using the full Eq.~(\ref{inseq2}) at a single temperature $T$, but with $S_{\rm bgd}$ at $\omega<0$ known from the low-$T$ step. Our approach, on the other hand, uses the full Eq.~(\ref{inseq2}) for all temperatures when fitting for $S_{\rm bgd}$. Thus our approach removes the need to have a measurement at extremely low temperatures in order to obtain an accurate estimate of the background contribution. The determining factor for the efficiency of our approach becomes the number and/or range of different temperatures, not their extreme values, enabling it to be used as a reliable procedure for background determination even when the range of measured temperatures is modest and no millikelvin measurements are available. By using all of the available data in one simultaneous fit, taking into account both $\omega<0$ and $\omega>0$ parts from measurements at all temperatures, our procedure also reduces the uncertainty in the determined background contribution $S_{\rm bgd}$ by an expected factor of $\sqrt{N}$ when compared to the approach of Refs.~\onlinecite{helton2007spin,helton2010dynamic}; which means a reduction in uncertainty by a factor of $\sqrt{2}$, even if we were to perform measurements at only two temperatures, the minimum for either of these procedures to work.

The above described novel background-determination procedure was used to obtain the background contribution to the INS measurements of \CuD, as shown in Fig.~\ref{fig4s}, and to subtract it from the total neutron-scattering intensity, determining the sample contribution [Fig.~3(a) in the main text] and its corresponding dynamical susceptibility [Fig.~3(b) in the main text]. The data sets were collected on the FOCUS time-of-flight spectrometer at the Paul Scherrer Institute, Switzerland, with an incident neutron energy of $5.50$~meV. Detector calibration was performed using a separate vanadium measurement.

\end{document}